# How much of the extreme luminosity of IRAS F10214+4724 can be attributed to gravitational lensing?

Neil Trentham

Institute for Astronomy, University of Hawaii

2680 Woodlawn Drive, Honolulu HI 96822, U. S. A.

email : nat@newton.ifa.hawaii.edu

# ABSTRACT

The galaxy IRAS F10214+4724, discovered in a spectroscopic survey of a 0.2 Jy sample by Rowan-Robinson and collaborators in 1991, is significantly more luminous than any other known galaxy. Its bolometric luminosity is $2 \times 10^{14}$ $L_\odot$, which is comparable to the luminosities of the most luminous quasars. Recent observations have revealed a candidate foreground group of galaxies, which might gravitationally lens F10214+4724, thus explaining much of its luminosity. High-resolution imaging of F10214+4724 has revealed that most of its near-IR flux comes from a circularly symmetric arc; this also supports the gravitational lens interpretation. In such a scenario, F10214+4724 would be the high-redshift analogue of the ultraluminous IRAS galaxies observed locally. This work presents a simple statistical lensing model to investigate this possibility.

We show that, on statistical grounds alone, the probability that F10214+4724 is a gravitational lens system with magnification $2 < \mu < 10$ is approximately 25%, if nearby determinations of the luminosity function $\phi(L)$ for ultraluminous IRAS galaxies can be extrapolated to both high redshifts $z$ and high luminosities $L$. If $\phi(L)$ steepens either with increasing $z$ or increasing $L$, we predict a substantial increase in the probability of F10214+4724 being a lens system. Very large magnifications ($\mu > 20$) are ruled out by this model, unless $\phi(L)$ is very steep, e.g. a power law with index $\alpha < -6$. These results therefore suggest that F10214+4724 is indeed the most luminous galaxy known. However, if it is a lens system with $\mu > 2$, it would not have been discovered had it not been lensed.

**Key words:** galaxies: individual: IRAS F10214+4724 – cosmology: gravitational lensing

## 1. INTRODUCTION

The identification of the IRAS Faint Source Catalog object F10214+4724 (hereafter F10214) as a galaxy at redshift $z = 2.286$ (Rowan-Robinson et al. 1991) has revealed an object of overwhelming bolometric luminosity ($2 \times 10^{14}$ $L_\odot$, for a Hubble parameter $H_0 = 75$ km s$^{-1}$ Mpc$^{-1}$ and cosmological density $\Omega_0 = 1$). This is higher than the luminosities of any other known galaxies and is comparable to the luminosities of the most luminous quasars. Subsequently the detection of CO 3−2 emission from this galaxy (Brown & Vanden Bout 1991) showed that is possesses enormous amounts of molecular gas ($\geq 10^{11} M_\odot$, Solomon, Radford & Downes 1992). This much gas suggests that the fraction of the dynamical mass of this galaxy that is in molecular gas is close to 100% (Solomon, Downes & Radford 1992a). It also suggests that F10214 might be a hyperluminous analogue of the ultraluminous infrared galaxies (ULIGs) observed at lower redshifts (Sanders et al. 1988; Solomon, Downes & Radford 1992b; Kim & Sanders 1995 hereafter KS95). In this paper, we present a simple model to investigate the suggestion that F10214 is a background galaxy that is gravitationally lensed by the foreground group of galaxies recently identified there (Elston et al. 1994).

Studies of gravitational lensing of distant objects can be broadly divided into two categories. The first of these is the observational study of specific objects. This includes studies of multiple images, distortions, and arcs, and the identification of candidate foreground objects. The second of these approaches is essentially statistical, and is based on the assertion that if matter is distributed inhomogenously on cosmological scales (i.e. in galaxies), then some fraction of objects at cosmological distances will have their observed fluxes amplified as their light is gravitationally deflected by these inhomogeneities. If we assume a distribution of matter, we can then calculate a probability spectrum for such amplifications and subtract it from observed luminosity functions, thereby obtaining intrinsic luminosity functions. This has been done extensively in studies of quasar source counts, but the observed galaxy luminosity function is only known in detail at low redshift ($z < 0.5$), where this effect is generally unimportant. This paper follows the second of these approaches;

we are forced to use the statistical approach because the direct observations are sufficiently inconclusive at present to address the question as to whether or not there is a significant lensing amplification. Whilst recognizing that in this case we are applying this statistical analysis to only one object, we attempt to calculate how probable it is that an object with the properties of F10214 at $z = 2.286$ is an object whose flux has been amplified by lensing relative to how probable it is that the object is unlensed and that the observed luminosity is the intrinsic one. The results are presented as functions of the lensing amplification and of the ULIG luminosity function at $z = 2.286$. In Section 2 we outline in detail why we might expect F10214 to be a gravitational lens system. In Section 3 we describe the calculations and our results. In Section 4 we discuss these results in the light of recent observational discoveries, and in Section 5 we summarize.

## 2. MOTIVATION

We consider first the observations which provide direct evidence supporting the hypothesis that F10214 is a background object that is gravitationally lensed by a foreground galaxy or group of galaxies. We then review the other observations of F10214 and discuss how such a scenario would affect our interpretations of these observations and how this supports (or refutes) the case for F10214 being lensed.

### a. Direct Evidence

In their near-infrared study of F10214, Elston et al. (1994) discovered several companions which they note have physical properties more characteristic of a foreground group of galaxies than of objects physically associated with F10214; they then point out that these galaxies may gravitationally lens F10214, hence explaining its large luminosity. Their results suggest that if this is indeed a foreground group, and if it has the total luminosity of an $L^*$ galaxy, it then has redshift $0.6 < z < 1$. They also note that the colours of the foreground group objects are consistent with those of early-type galaxies at this redshift. This redshift range corresponds to distances approximately half that to F10214, and so is optimal for the redshift of lenses yielding large lensing amplifications. This is therefore a primary motivation for the work presented here.

In addition, Matthews et al. (1994) observed F10214 with high resolution using the Near-Infrared Camera (NIRC) on the 10-metre W. M. Keck Telescope. Their deconvolved image revealed that a substantial fraction of the observed near-IR flux comes from a structure similar to an arc such as those produced by gravitational lensing. This result has since been spectacularly confirmed by Graham & Liu (1995), who have also imaged F10214 with NIRC on the Keck Telescope, but with higher resolution (FWHM of $0.4''$).

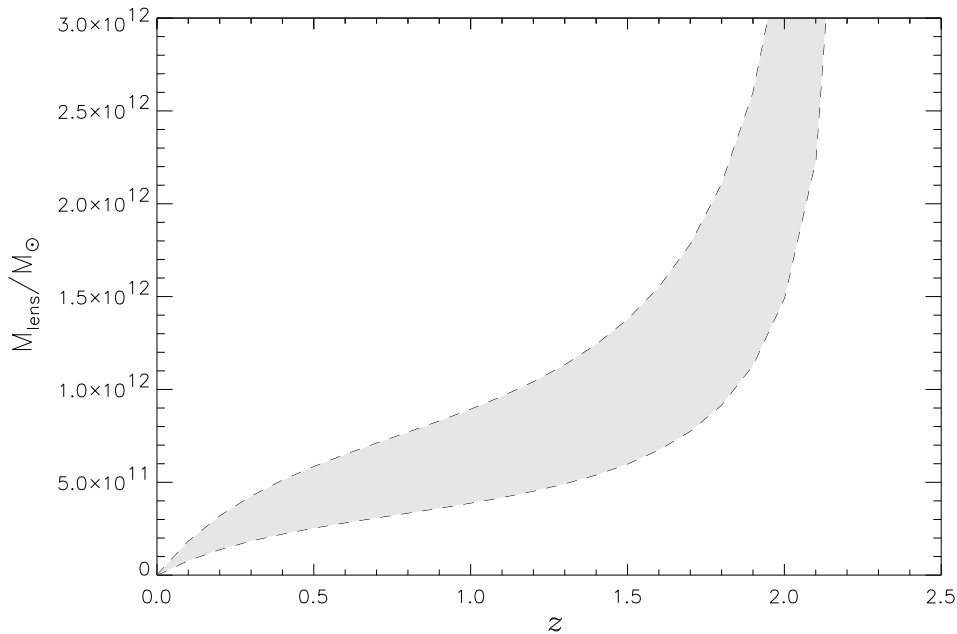

**Figure 1.** Combinations of lens mass $M_{\rm lens}$ and redshift $z$ permitted if Source 1 of Matthews et al. (1994) is to be interpreted as an arc resulting from gravitational lensing by Source 2. Source 1 is then assumed to be at $z = 2.286$. Here, as throughout this paper, we assume $H_0 = 75$ km s$^{-1}$ Mpc$^{-1}$, and $\Omega_0 = 1$.

The angular radius $\theta$ of the arc observed by Matthews et al. (1994) is $1.4'' < \theta < 2.1''$. For an Einstein ring,

$$\theta = \left(\frac{4GD_{\text{ls}}M_{\text{lens}}}{c^2 D_{\text{s}} D_{\text{l}}}\right)^{\frac{1}{2}}, \tag{1}$$

where $c$ is the speed of light, $G$ is Newton's constant and $M_{\text{lens}}$ is the lens mass (assumed spherically symmetric) interior to the arc. Here $D_{\text{ls}}$, $D_{\text{s}}$, and $D_{\text{l}}$ are the lens-source, observer-source, and observer-lens angular diameter distances, respectively. Therefore the values of $\theta$ that Matthews and collaborators observe are consistent with the value of $\theta$ which would be produced were an object with the mass of a large galaxy or small group of galaxies at redshift $0.6 < z < 1$ to lens a compact background object at redshift 2.286 (Fig. 1).

b. **Indirect Evidence**

Assuming that the observed luminosity of F10214 at all wavelengths comes from the same compact region, we can assume that the lensing process is achromatic and that the observed spectral energy distribtion (SED) of F10214 (Fig. 2) is simply a constant multiplied by the intrinsic SED. Figure 2 suggests that F10214 has a similar shape SED to Markarian 231 (Mrk 231) and Arp 220, but is $50 - 100$ times more luminous. The similarity in SEDs suggests that F10214 is a more luminous example of the class of objects containing Mrk 231 and Arp 220, that is the ULIGs (Sanders et al. 1988). Gravitational lensing would then bring these curves closer together. The least secure aspect of this interpretation is that the X-ray luminosity of F10214 (Lawrence et al. 1994) is somewhat higher than 100 times the Mrk 231 upper limit and the Arp 220 detection, so that the amplification at short wavelengths appears to be achromatic. However, the work of Eales & Arnaud (1988) suggests that there may be huge uncertainties in the X-ray extinction (and hence intrinsic X-ray luminosity) of galaxies that are known to contain large quantities of dust; this may explain the inconsistency. Furthermore the X-ray detection of F10214 is only at the $2\sigma$ level (Lawrence et al. 1994), so its true X-ray flux might be somewhat lower than these authors quote. The direction of polarization of light is unaffected by gravitational lensing (Dyer & Shaver 1992), so we are unable to use the available polarimetric data on F10214 (Lawrence et al. 1993, Januzzi et al. 1994) to obtain information about a potential amplification. Finally, lensing would make the mass of $H_2$ relative to the dynamical mass fall substantially below 100% by reducing $L_{CO}$ and therefore the $H_2$ mass (but not the CO line velocity and therefore not the dynamical mass; see Solomon, Downes & Radford 1992a). In summary, moderate lensing amplifications ($\sim 5$) would suggest that the fraction of the dynamical mass which is in $H_2$ is $\sim 30\%$ (using the results of Solomon, Downes & Radford 1992a). Magnifications substantially higher than this would result in an anomalously low fraction of molecular gas. In such a scenario, F10214 would then be about 10 times more luminous than Mrk 231. It would still be the most luminous galaxy known, but it would have easily escaped detection in the IRAS Faint Source Catalog had it not been lensed.

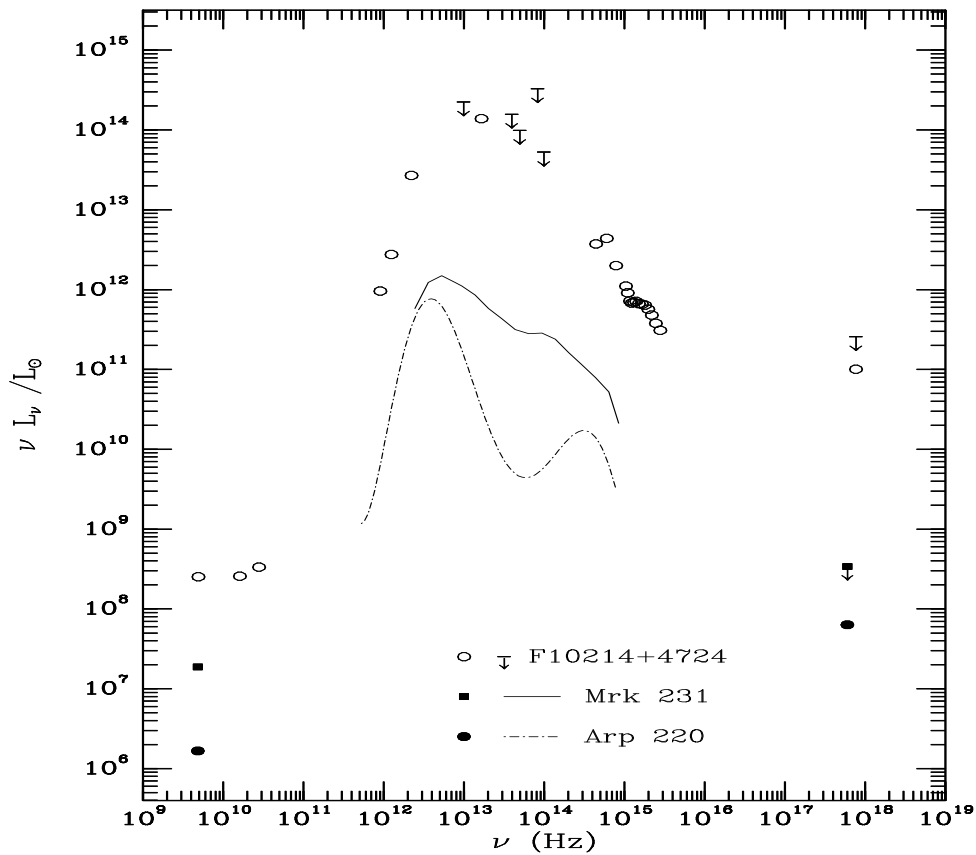

**Figure 2.** The rest-frame spectral energy distribution (SED) of F10214 (Rowan-Robinson et al. 1993, Lawrence et al. 1994). The SEDs of Markarian 231 and Arp 220 (Sanders et al. 1988, Condon & Broderick 1991, David et al. 1992) are also shown.

## 3. MODEL AND RESULTS

Two distinct approaches, one geometrical and one statistical, suggest themselves to further investigation of the possibility of F10214 being a gravitational lens system with significant magnification. The most natural approach is to construct specific models assuming some input lensing parameters and to compare these models to the observed morphology of F10214. An example of an object where this approach has been very successful is the Cloverleaf quasar (Kayser et al. 1990), which has now also been detected in CO (Barvainis et al. 1994). However, we do not follow this approach for F10214, for two reasons. Firstly, the observational constraints on the morphology of F10214 on small angular scales are too weak to rigorously constrain such models. Secondly, the translation from an observed image to the numerical value of the magnification through this kind of modelling is sufficiently non-unique that even if very high-resolution data was available, it is not clear that we would be able to calculate the total magnification and so answer the question posed in the title of this paper. For example, in the case of the Cloverleaf quasar, where there is abundant multiple-image and time-delay information available, there is substantial uncertainty in the value of the total amplification, even though models exist which fit the data extremely well (Kayser et al. 1990). We therefore are forced to adopt the statistical approach which is outlined in this section.

A comprehensive treatment of the theory of statistical gravitational lensing is given by Schneider, Ehlers, and Falco (1992, hereafter SEF92). In developing our model, we shall follow their notation, and quote some of the basic results here. For the detailed derivations the reader is referred to SEF92. In outline, our strategy will be to compute the magnification probability of extended sources (specifically that of uniform circular disks with sizes characteristic of F10214) for point-mass lenses assuming, some properties of the lenses and some cosmology, and then to correct this probability spectrum to allow for the extended nature of the lenses using the results of Kayser & Refsdal (1988). We will then combine this corrected probability spectrum with an assumed hypothesized luminosity function of ULIGs at $z = 2.286$ in order to determine the probability (as a function of the magnification $\mu$) that a galaxy with the properties of F10214 is magnified by $\mu$ relative to the probability of its being unlensed. This will then be done for a series of hypothesized luminosity functions.

The probability $P(>\mu_e)$ of an extended source at redshift $z_s$ being magnified more than $\mu_e$ through gravitational lensing from a randomly positioned source along our line of sight is equal to (SEF92 Eq. 12.41a)

$$P(>\mu_e) = \frac{3}{2}\Omega_L <\mu>(z_s) \int_0^{z_s} \frac{r(z)r(z,z_s)(1+z)dz}{<\mu>(z)r(z_s)\sqrt{1+\Omega_0 z}} \int_0^\infty m\,dm\,N(m)y^2(\mu_e, R(z,m)). \tag{2}$$

The notation is as follows: $\mu_e$ is the scalar magnification, equal to determinant of the inverse of the Jacobian matrix of the mapping from the image plane to the source plane; $\Omega_L$ is the cosmological density of lenses, in units of the critical density; $\Omega_0$ is the total cosmological density (assuming a matter-dominated universe); $r(z_1, z_2)$ and $r(z) = r(0, z)$ are the solutions to the Dyer-Roeder (1973) equation, which describes light propagation in an inhomogeneous universe (see SEF92 Section 4.5.3) given $\Omega_L$ and $\Omega_0$; $m$ is the mass of an individual lens in units of some reference mass $M_0$, and $N(m)$ is the number density of these lenses normalized such that $\int_0^\infty N(m)m\,dm = 1$; $<\mu>(z)$ is the angular average magnification, normalized to be 1 for a universe in which all matter is smoothly distributed (SEF92 Eq. 4.82); $R(z,m)$ represents the dimensionless source size (SEF92 Eq. 12.41b,c)

$$R(z,m) = R_0 \sqrt{\frac{r(z)}{mr(z_s)r(z,z_s)}}, \tag{3}$$

where

$$R_0 = \Gamma\sqrt{\frac{cH_0}{4GM_0}}, \tag{4}$$

and $\Gamma$ is the physical source size; $y(\mu_e, R)$ represents the source position, and, assuming that the source is a uniform circular disk, is given by the inverse function of (SEF92 Eq. 11.12)

$$\mu_e(y, R) = \frac{2}{\pi R^2} \int_{|y-R|}^{y+R} dr \frac{r^2+2}{\sqrt{r^2+4}} \arccos\frac{y^2+r^2-R^2}{2yr} + \mathrm{H}(R-y)\frac{R-y}{R^2}\sqrt{(R-y)^2+4}, \tag{5}$$

where H($x$) is the Heaviside step function. We consider in detail the case when all the lenses have the same mass $M_0$ i.e. $N(m) = \delta(m-1)$, and are distributed homogeneously. This is a reasonable approximation because we shall ultimately consider the case where the lensing objects are normal giant galaxies, and for a distribution of galaxies that obeys a Schechter (1976) luminosity function (assuming a constant global mass-to-light ratio for giant galaxies), most of the mass comes from objects within a very narrow mass range. For example, if the faint-end slope of the Schechter function is close to $-1$ (Davis & Huchra 1982), half the mass comes from objects with masses within a factor of two of that of an $L^*$ galaxy, where $L^*$ is the characteristic luminosity of the Schechter function; if the luminosity function describing giant galaxies is not Schechter but Gaussian, as is suggested by the observations of local groups by Ferguson & Sandage (1991), then this constraint on the mass range of lenses is tighter still. Making this assumption about the mass distribution of lenses, we can calculate $P(>\mu_e)$ from Equation (2). The results are displayed in Fig. 3, for a number of different cosmologies and values of $R_0$ (see Table 1). For a source at $z = 2.286$,

$$R_0 = 0.051 \left(\frac{\Gamma}{1\text{kpc}}\right) \left(\frac{M_0}{5 \times 10^{11} M_\odot}\right)^{-0.5} h_{75}^{0.5}, \qquad (6)$$

where $H_0 = 75\, h_{75}$ km s$^{-1}$ Mpc$^{-1}$. The case $R_0 = 0$ corresponds to point-sources. Figure 3 suggests that at very high magnifications $P(>\mu_e) \sim \mu_e^{-6}$ (see also SEF92 Section 12.5.1), so that these extreme magnifications are extremely rare. For intermediate magnifications $P(>\mu_e) \sim \mu_e^{-2}$. Given $R_0$, there exists some critical magnification $\mu_c$ that marks the transition between these two regimes; $\mu_c$ decreases with increasing $R_0$. The dependance of $P(>\mu_e)$ on the cosmology is two-fold: $P(>\mu_e)$ depends on (i) the number of lenses per line of sight and hence on $\Omega_L$, and (ii) the variation of the comoving volume element with redshift. The second of these effects arises from the dependance of the solutions of the Dyer-Roeder equations on the cosmological parameters and is significantly the weaker of these two effects. This means that in order to investigate a model in detail, it is important to choose $\Omega_L$ accurately, and our choice of the total density $\Omega_0$ is less critical (at higher source

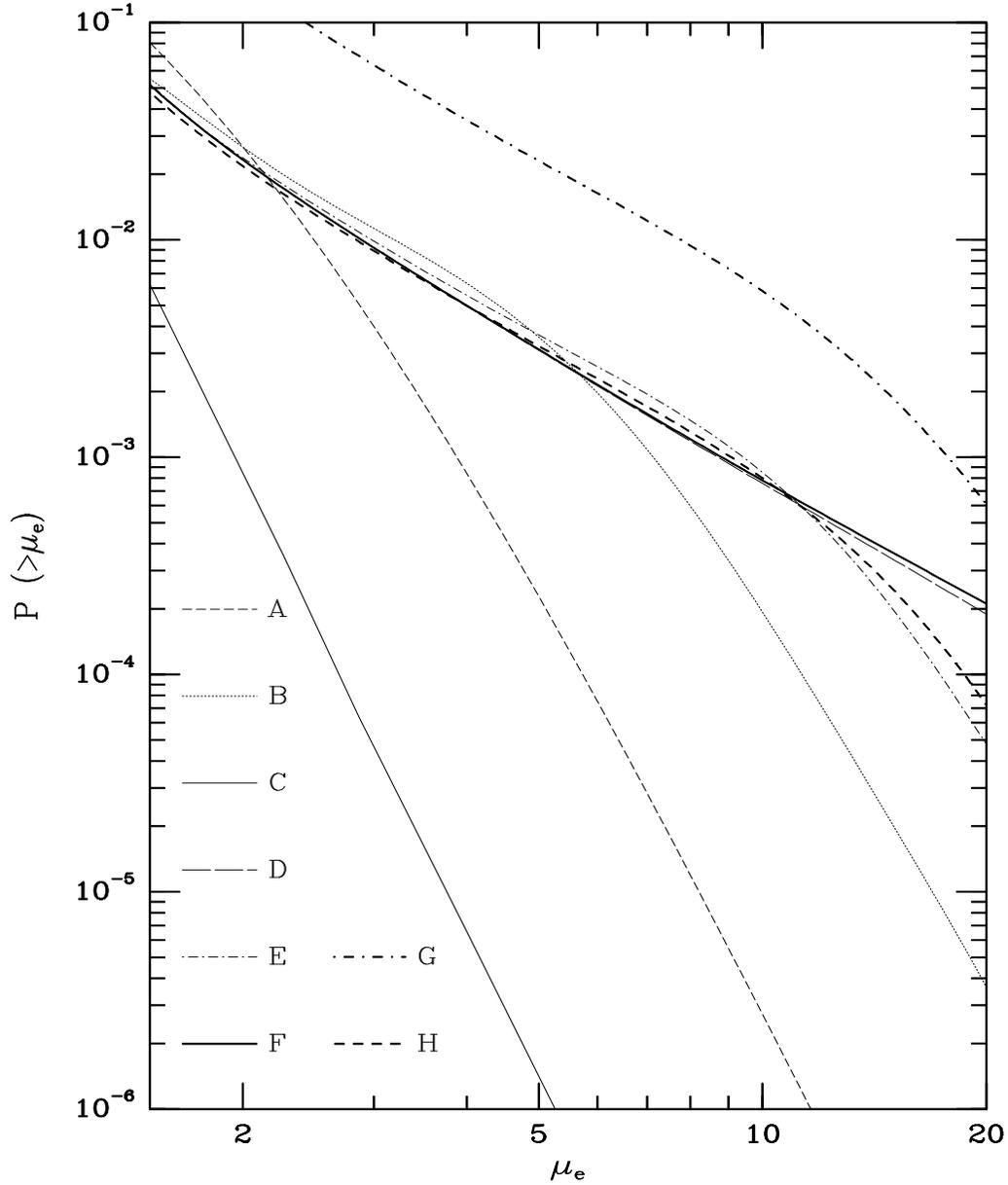

**Figure 3.** The probability $P(>\mu_e)$ that an extended source is magnified by more than $\mu_e$. The source is at redshift 2.286 and is assumed to be a uniform brightness circular disk with dimensionless radius $R_0$. The lenses are assumed to be point-like, of constant mass, and distrubted uniformly with cosmological density $\Omega_L$ in the present epoch. The cosmological matter density is $\Omega_0$ in units of the critical density. Curves for each of the eight models outlined in Table 1 are shown.

## Table 1: Models

| Model | $R_0$ | $\Omega_0$ | $\Omega_L$ |
|-------|-------|------------|------------|
| A | 0.3 | 1 | 0.2 |
| B | 0.1 | 1 | 0.2 |
| C | 1.0 | 1 | 0.2 |
| D | 0 | 1 | 0.2 |
| E | 0.05 | 1 | 0.2 |
| F | 0.01 | 1 | 0.2 |
| G | 0.05 | 1 | 1 |
| H | 0.05 | 0.2 | 0.2 |

redshifts, the reverse becomes true, e.g. Turner 1990). These conclusions about the cosmology have assumed a matter-dominated universe with a vanishing cosmological constant $\Lambda$. Although the Dyer-Roeder equations are not valid in a universe where $\Lambda \neq 0$, the effects of a non-zero $\Lambda$ term can be qualitatively investigated using the Turner-Ostriker-Gott (1984) optical depth ($\tau$) formulation. We note a change in optical depth corresponding to $\Delta \mathrm{Log}_{10} \tau$ ($\approx \Delta \mathrm{Log}_{10} P(> \mu_e)$) of approximately 0.5 on going from an $\Omega = 1$ matter dominated Universe to an $\Omega_{\mathrm{matter}} = 1 - \frac{\Lambda}{3 H_0^2} = 0.3$ Universe (that favored by large-scale structure observations and the solution to the cosmological timing problem; Kofman et al. 1993) if $\Omega_L$ is kept the same. This is smaller than the difference between Models E and G (where the only difference is the value of $\Omega_L$) in Fig. 3, but is significant nonetheless.

We will further consider Model E in detail. This model further adopts a value of $\Omega_L$ which is the near the upper end of the range consistent with observations (Trimble 1987, Tully et al. 1993). Furthermore, if all the mass in lenses is contained within isothermal spheres of masses between $5 \times 10^{11}$ M$_\odot$ and $1 \times 10^{12}$ M$_\odot$ Model E results in a lensing $F$ parameter of $0.1 < F < 0.2$ (here $F$ is a dimensionless parameter describing the lensing effectiveness of a population of singular isothermal sphere lenses; Turner et al. 1984), which is close to the value normally adopted in studies of the statistical lensing of quasars (Turner 1990, Fukugita & Turner 1991). Nevertheless the choice of $\Omega_L$ is the biggest uncertainty in this approach. However, $P(>\mu_e)$ scales simply as $P(>\mu_e) \sim \Omega_L^{1.5}$ for $\Omega_L < 1$ so that this effect may be quantified straightforwardly (the relationship is not linear as might be suggested from Equation (2), because both $<\mu>$ and the solutions to the Dyer-Roeder equations are functions of the ratio $\frac{\Omega_L}{\Omega_0}$). Figure 4 shows the dependance of the $P(>\mu_e)$ profile on the source redshift, assuming Model E. The figure suggests that significant magnifications ($\mu_e > 2$) are only important for high-redshift sources; the effects of lensing on the local galaxy luminosity function, which is only known in detail at redshifts $z < 0.05$, are negligible. This rapid increase in lensing probability with redshift is well known; Peacock (1982) has shown that the optical depth to lensing $\tau \sim z^3$ at small $z$. Figure 5 shows for Model E with the source at $z = 2.286$ the conditional probability spectrum of the lens redshift given that the source has been magnified by a known amount. The figure suggests that for intermediate magnifications ($\mu \sim 5$), lenses that are at very low redshifts or that are very close to the source are unlikely.

The derivation of Fig. 3 has assumed that the lenses are point masses. If, however, we assume that individual galaxies are the lensing objects, we need to construct a transfer function $T$ to correct for the extended nature of the lenses. We do this by assuming that the lenses are well approximated by isothermal spheres (Turner et al. 1984), and then (i) noting that the isothermal-sphere-to-point-mass cross-section ratio $q(\mu_e, z)$ (for explanation, see Fig. 6) is close to one (specifically that over the range of $\mu_e$ we are considering, the integral $\int_0^{z_s} g_{\mu_e \infty}(z) q(\mu_e, z) dz$ does not differ from unity by more than a few percent, where $g_{xy}(z)$ is as in Fig. 5), and (ii) adopting the model of Kayser & Refsdal (1988, hereafter KR88) which uses King galaxies as lenses, and setting $T = \frac{\kappa(\hat{f}=f_{\text{gal}})}{\kappa(\hat{f}=0)}$. Here $\kappa(\hat{f})$ is as defined

– 13 –

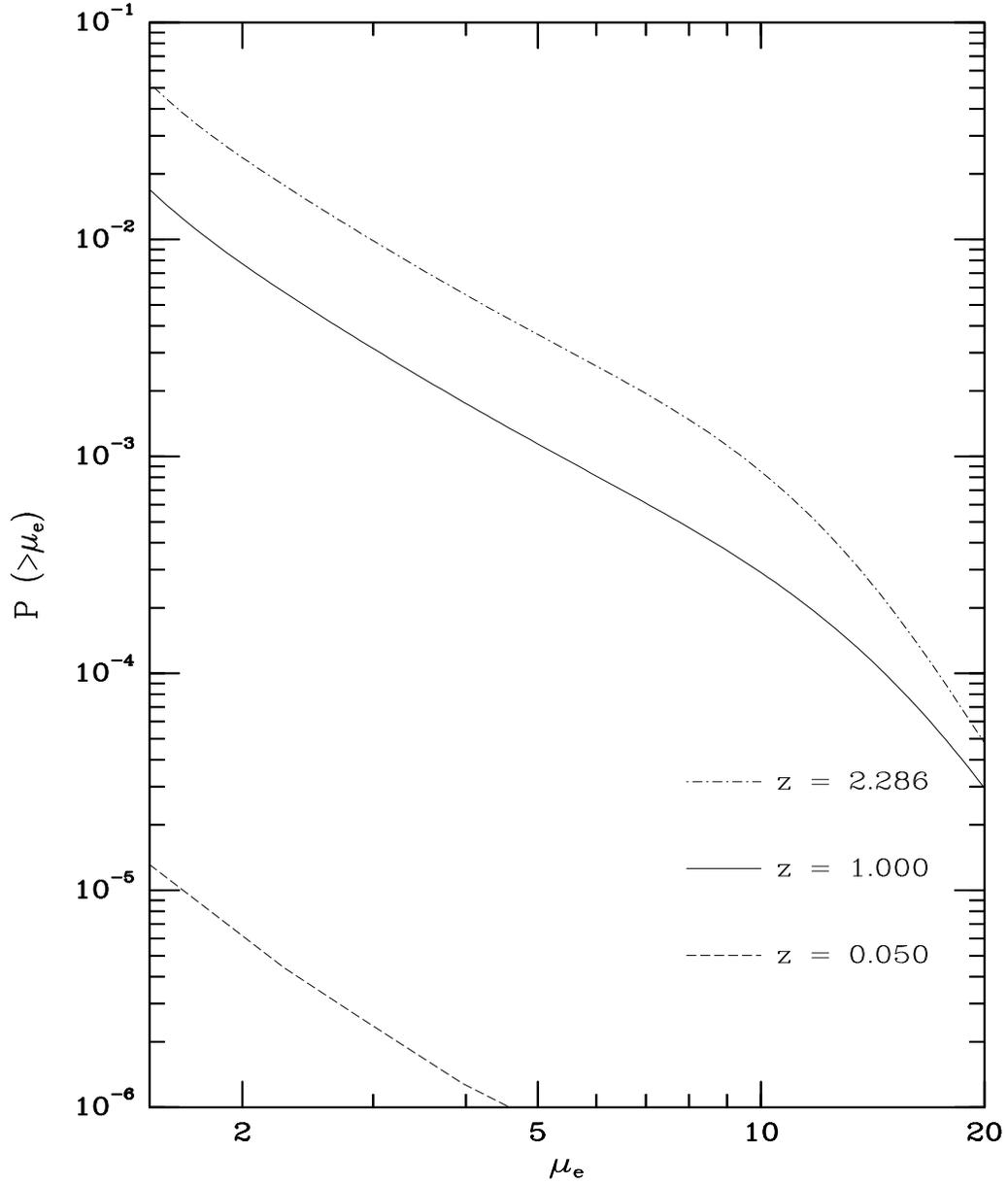

**Figure 4.** As Fig. 3, but for sources at three different redshifts. The parameters of Model E are assumed.

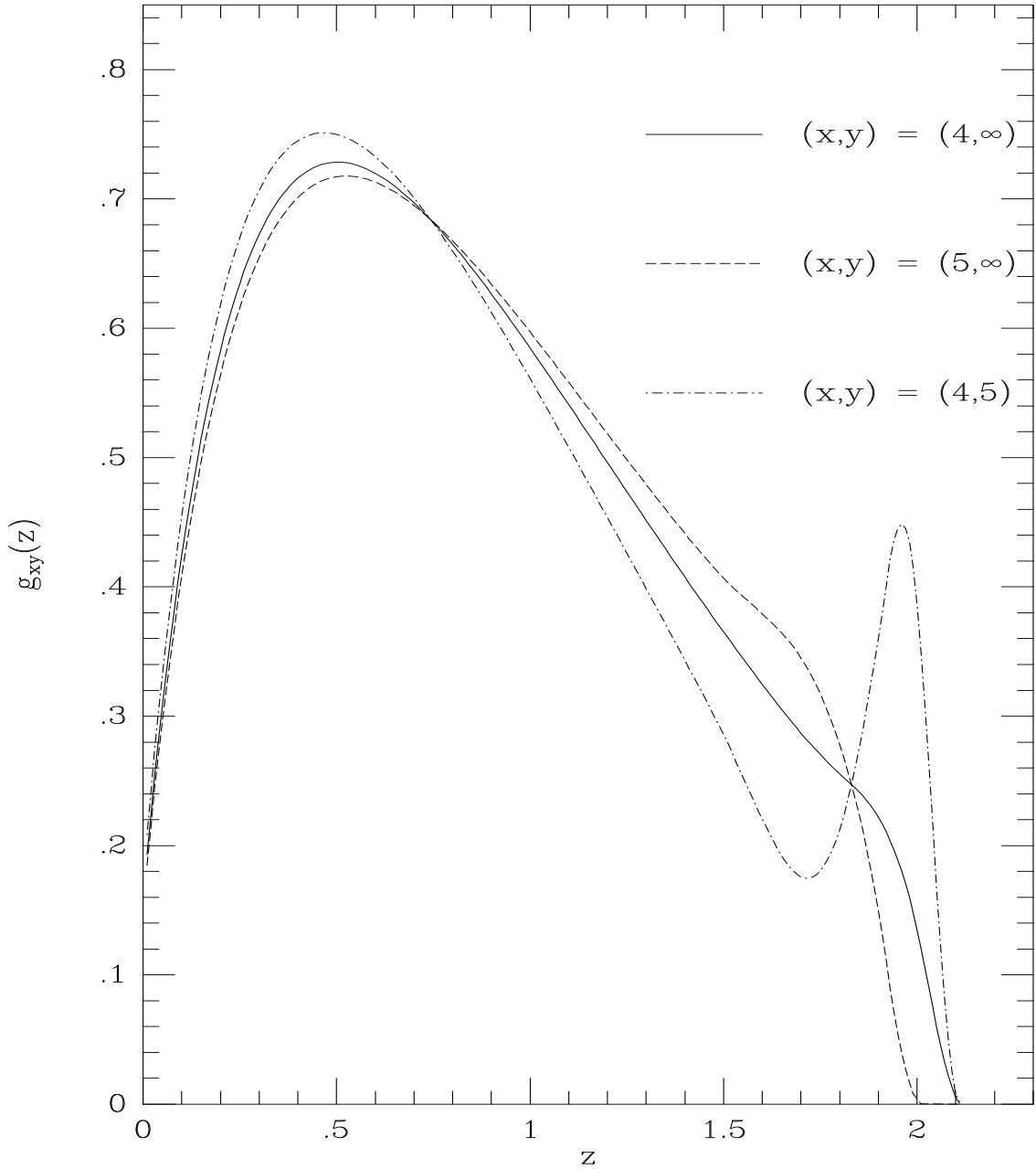

**Figure 5.** Examples of the function $g_{xy}$, where $g_{xy}dz$ is the conditional probability of the lens being between redshift $z$ and $z + dz$, given that lens has magnified the source by a factor $x < \mu_e < y$. The properties of the source and lens are the same as for Fig. 3, assuming Model E. Note that $\int_0^{z_s} g_{xy} dz = 1$ for all $(x, y)$.

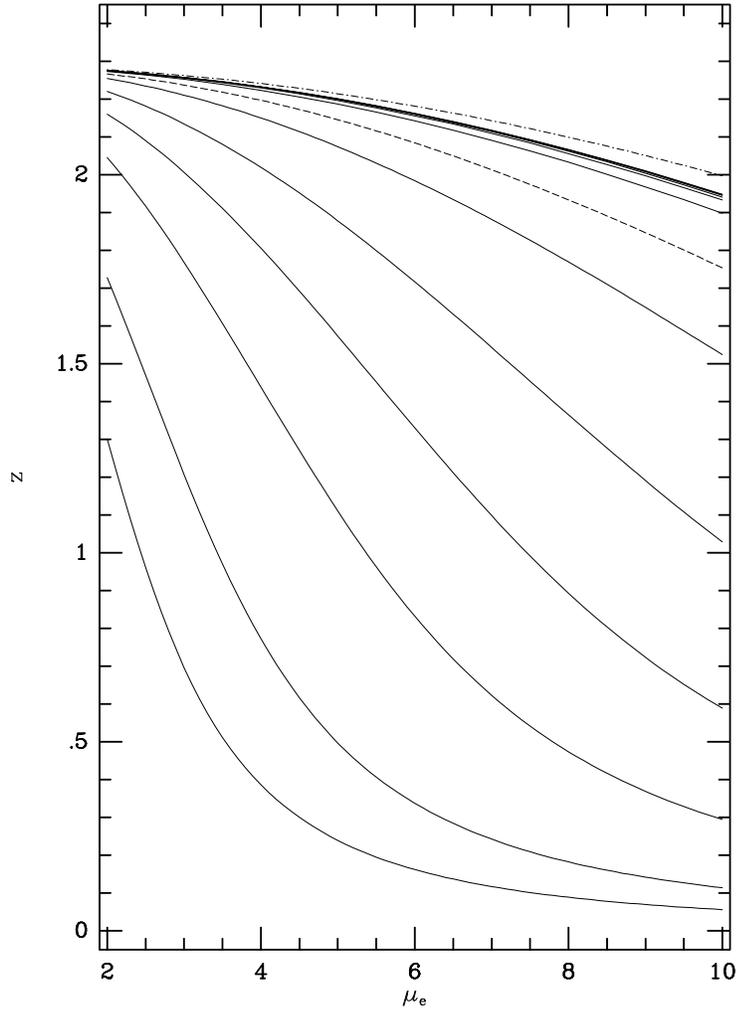

**Figure 6.** Contours of constant $q(\mu_e, z)$ where $q(\mu_e, z) = \frac{\sigma_e(\mu_e, z)}{\sigma_p(\mu_p, z)}$ is the ratio of the cross-section of an extended source of dimensionless size $R_0 = 0.05$ at $z = 2.286$ to lensing with magnification $> \mu_e$ from an isothermal sphere at redshift $z$ to the cross-section of a point source at $z = 2.286$ to lensing with magnification $> \mu_p \equiv \mu_e$ from the same isothermal sphere. The dotted line represents the maximum value of $q = 1.80$ (given by the solution of the simultaneous equations $\zeta\left(\frac{\sqrt{q}}{b_0}\right) = b_0$ and $\left|\frac{\partial}{\partial b}\zeta\left(\frac{\sqrt{q}}{b}\right)\right|_{b=b_0} = 1$, where $\zeta(x)$ is given by Equation (9)). The contours on either side of this line represent $q = 1.50, 1.20, 1.10, 1.05, 1.02$, and $1.01$, respectively with increasing distance from the dotted line (note that the furthest four lines on the upper side are too close to be distinguishable). The dotted-dashed line represents the value $q = 0$. Above this line, extended sources cannot be magnified by more than $\mu_e$ by sources at that particular $z$. This limit follows from the condition $\zeta(x) \leq 2$.

by Equation (12) of KR88 for the case $D_c = 0.15$ (as is appropriate for a source at $z_s = 2.286$ assuming Model E) and

$$f_{\text{gal}} = 0.51 \left(\frac{\Xi}{20\text{kpc}}\right)^2 \left(\frac{M_0}{5 \times 10^{11} M_\odot}\right)^{-1} h_{75}^{-1}, \qquad (7)$$

where $\Xi$ is the lens size. We then adopt $f_{\text{gal}} \sim 0.9$ and so derive $T = 0.92$ from Equations (14a,b) of KR88. Equations (14a,b) in KR88 further suggest that $T$ is a weak fuction of $f_{\text{gal}}$, so that the exact choice of $f_{\text{gal}}$ is not critical. The transfer function is a constant because over the range of magnifications that we are considering, the profile $P(> \mu_e)$ has the same functional form (inverse square) for both pointlike and extended lenses (Fig. 3, SEF92 Section 11.4.1). Note that in the models of KR88, the lenses are approximately isothermal spheres so that in computing Fig. 6, we can assume that for such lenses, the cross section is given by $\sigma = \pi y^2$, where $y$ is the inverse of the function (see SEF92 Section 8.1.1)

$$\mu_e = \frac{2}{R} \zeta\left(\frac{y}{R}\right), \qquad (8)$$

and

$$\zeta(w) = \begin{cases} \frac{4}{\pi} \text{E}(w), & \text{if } w \leq 1; \\ \frac{4}{\pi} w \left(\text{E}\left(\frac{1}{w}\right) - (1 - \frac{1}{w^2})\text{K}\left(\frac{1}{w}\right)\right), & \text{if } w \geq 1, \end{cases} \qquad (9)$$

where $\text{K}(x)$ and $\text{E}(x)$ are the complete elliptic integrals of the first and second kind respectively.

We are now in a position to combine our probability spectrum $P(> \mu_e)$ and transfer function $T$ with an intrinsic ULIG luminosity function $\phi(L)$ at $z = 2.286$. We calculate the probability $\rho(\mu_e)$ that an object with observed luminosity $L$ at this redshift is magnified by an amount between $\mu_e$ and $\mu_e + \delta\mu_e$ and the corresponding probability $\rho(1)$ of it being unlensed or very weakly lensed. The results are then expressed as the ratio of these two quantities in the limit as $\delta\mu_e \to 0$. For a power-law $\phi(L)$, this ratio $\frac{\rho(\mu_e)}{\rho(1)}$ is given by the relation

$$\frac{\rho(\mu_e)}{\rho(1)} = C\mu_e^{-\alpha} P(> \mu_e) T. \qquad (10)$$

Here $\alpha = \frac{d\log\phi(L)}{d\log L}$ is the slope of $\phi(L)$, and $C$ is a correction factor that accounts for the steepening of the $P(>\mu_e)$ profile (Fig. 3) at low magnifications. Uncertainty in $T$ at these low magnifications (see Fig. 1 of KR88 for the case $\hat{f} = 0$ at $D = 0$ for $D_c \sim 0.1$) means that we cannot calculate this profile in detail in this range. Instead we approximate $P(>\mu_e)$ to be a power law in the range $1 < \mu_e < 1.5$ and so derive $C = 0.275$.

Expressing the result this way allows the result to be independant of the density evolution of $\phi(L)$, about which there is huge uncertainty (Saunders et al. 1990, Fisher et al. 1992). Figures 7 and 8 show $\frac{\rho(\mu_e)}{\rho(1)}$ for various scenarios. The value of $\alpha = -3.3$ is what is observed for the ULIGs in the highest redshift complete sample that has been studied to date (KS95). In this sample the mean redshift is 0.3 and luminosity is $10^{12.5}$ $L_\odot$. A larger $|\alpha|$ corresponds to a steepening of $\phi(L)$ at higher luminosities; such a steepening is observed for quasars (Giallongo & Vagnetti 1992). Figure 7 shows how more pronounced this effect is for ULIGs at $z \sim 2$ than for either ULIGs at low $z$ (due to the greater optical depth to lensing) or for QSOs at $z \sim 2$ (due to the QSOs having a shallower luminosity function at high luminosities: Marshall 1985, Giallongo & Vagnetti 1992). These figures are valid in the range $2 < \mu_e < 10$. We may, however, investigate the qualitative behaviour of $\frac{\rho(\mu_e)}{\rho(1)}$ on either side of this range. On the low $\mu_e$ side, $\frac{\rho(\mu_e)}{\rho(1)}$ turns over and increases, eventually becoming equal to 1 at $\mu_e = 1$. The turnover results from the complex behaviour of $P(>\mu_e)$ at $\mu_e \approx 1$ and the resultant steepening from a $\mu_e^{-2}$ profile (see Fig. 12.13 in SEF92). The details of the profile in this range will depend on the details of the transfer function, which are uncertain (see above). On the high $\mu_e$ side $\frac{\rho(\mu_e)}{\rho(1)}$ reaches a maximum near the turnover $\mu_e$ shown in Fig. 3 (this is about 10 for Model E) and then drops rapidly due to the $\mu_e^{-6}$ tail.

As is stressed at the beginning of this section, specific models of F10214 are not presented as these are not sufficiently unique to constrain $\mu_e$, especially given the resolution of the present observations. Nonetheless, it is an important plausibility check to ensure that models exist which simultaneously provide significant magnification and explain any morphological peculiarities of F10214 if gravitational lensing is to be invoked as an explanation of its strange properties. This will be an valuable exercise when high-resolution data become available for this object. An equally important plausibility check is that models exist that are simultaneously consistent with the near-infrared, radio, and CO observations, given a plausible radial distribution of CO and stars in the background galaxy. There is already some evidence for this in that the position angles of the extended emission in the near-infrared (Matthews et al. 1994), CO (Radford et al. 1993) and 1.49 GHz (Lawrence et al. 1993) maps are all consistent with each other (at approximately N88°E), but, again, higher resolution information is required for this exercise to be rigorous.

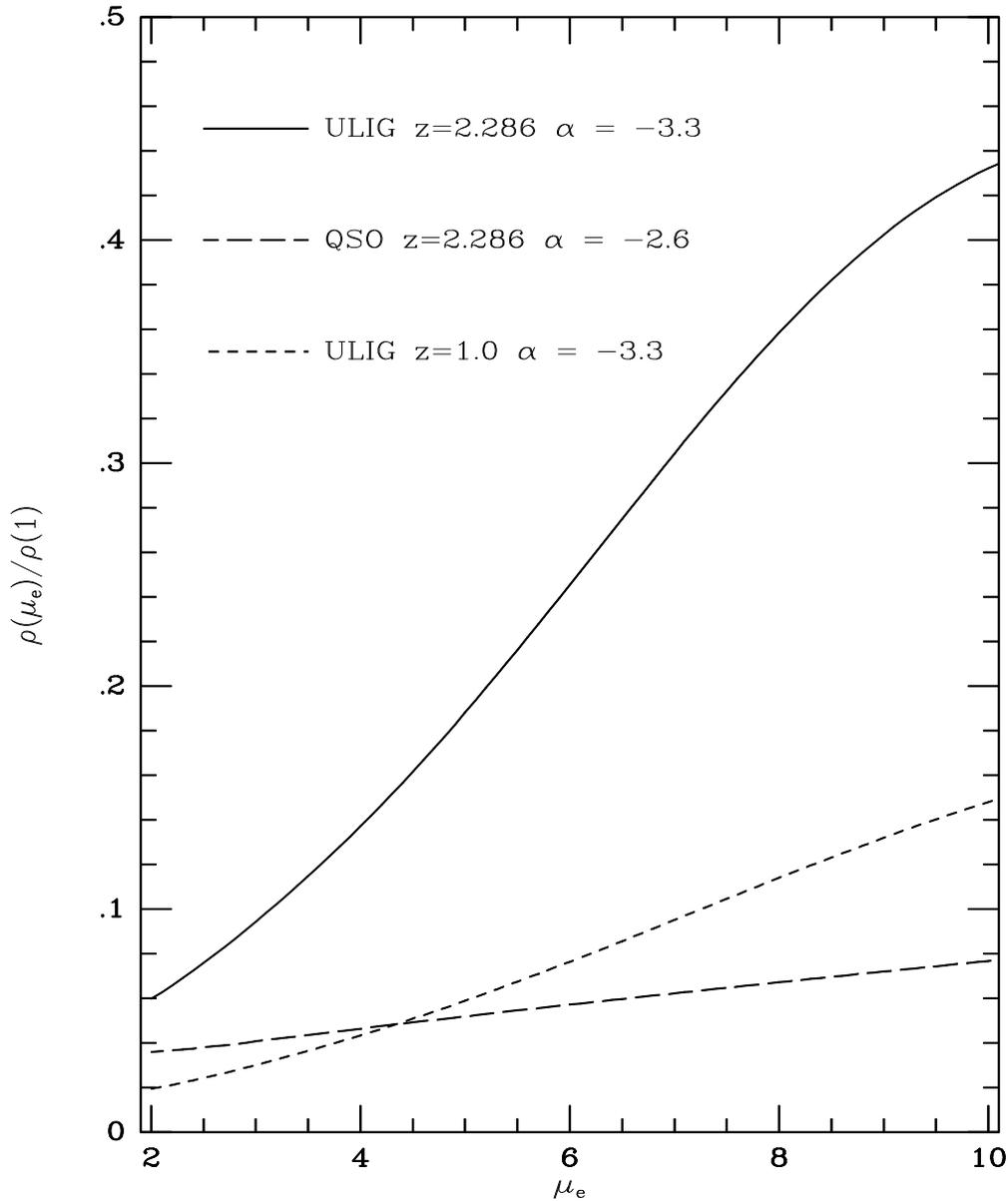

**Figure 7.** The probability $\frac{\rho(\mu_e)}{\rho(1)}$ of an object that we observe at redshift $z$ being gravitationally lensed by magnification $\mu_e$ relative to the probability of it being unlensed. Here $\alpha = \frac{d\log\phi(L)}{d\log L}$ is the luminosity function slope for the class of objects being considered at redshift $z$, which is assumed to be a power-law for the luminosity range $\frac{L_0}{\mu_e} < L < L_0$, where $L_0$ is the observed luminosity of F10214. The cosmology described by Model E is assumed, along with $R_0 = 0.05$ (ULIGs) or 0 (QSOs), and the transfer function of Kayser & Refsdal (1988) has been adopted.

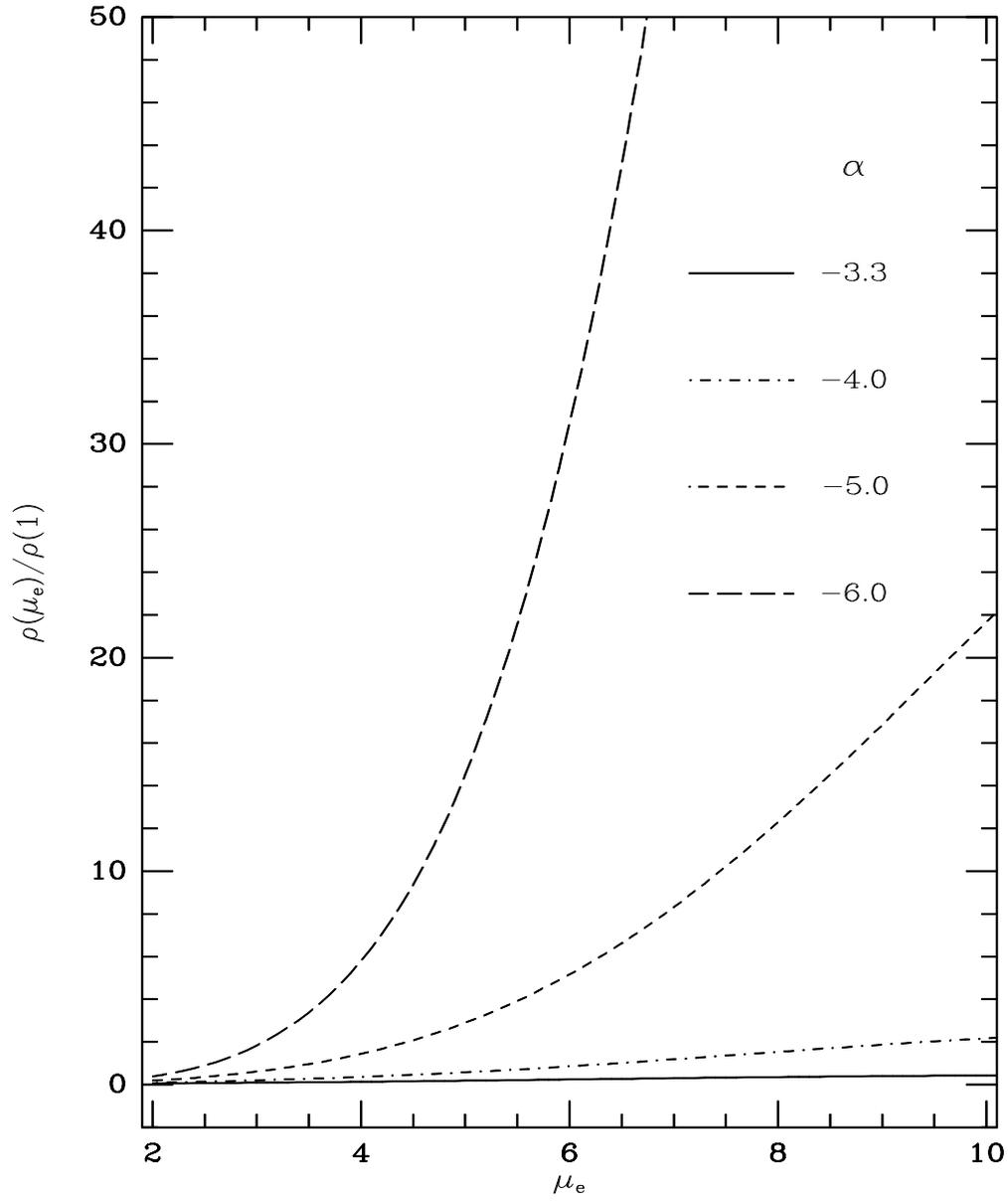

**Figure 8.** As Fig. 7, but for ULIGs at $z = 2.286$ with various $\alpha$. The value $\alpha = -3.3$ is what is observed locally (KS 95).

## 4. DISCUSSION

In this section we discuss the implications of the results of the previous section. First we examine the various model parameters, the observational constraints on these parameters, and how varying these parameters affects the case for F10214 being lensed. We then discuss possible observational tests of this model. Finally, we consider the implications for the studies of cosmology and of ULIGs at high redshift should this model be a realistic description of F10214.

Figure 7 implies that there is an approximate probability of 25% of an object with the properties of F10214 being a gravitational lens system with magnification $2 < \mu_e < 10$ assuming the cosmology and lens properties described by Model E, and assuming that the ULIG luminosity function of KS95 can be extended to higher luminosities and higher redshifts. Magnifications much steeper than this are ruled out by this model due to the $\mu_e^{-6}$ tail at high magnifications. A steepening of the luminosity function results in a rapid increase in the probability of F10214 being lensed. Such a steepening is observed when the luminosity functions of (i) normal galaxies in both clusters and the field (Felten 1985), and (ii) quasars have been extended to higher luminosities; no such steepening has yet been observed in the luminosity function of ULIGs. However, these luminosity functions only extend to luminosities less than $\frac{1}{10}$ of the observed luminosity of F10214. Furthermore, the observed luminosity function of ULIGs has only been determined out to $z = 0.3$; if galaxies form by hierarchical clustering, we might expect the slope of the bright-end of the luminosity function to be steepen with increasing $z$. For comparison, the slope of the luminosity function required for the probability of F10214 being lensed with $2 < \mu_e < 10$ to be greater than 95% corresponds to the slope of the Schechter (1976) luminosity function at 7 $L^*$.

We shall now examine the assumptions (other than the value of $\alpha$) inherent in this model. In particular, we have adopted values of $\Omega_L$ and the lens mass $M_0$ (and therefore $R_0$) that, whilst being consistent with observations, tend to lead to high values of the lensing $F$ parameter. We shall examine each assumption in turn and then consider how much larger (or smaller) $|\alpha|$ need be for $\frac{\rho(\mu_e)}{\rho(1)}$ to remain constant if we vary $\Omega_L$ and $M_0$.

(i) Source geometry: we have assumed that the source is a uniform circular disk in the plane of the sky with physical radius $\Gamma \sim 1$ kpc. If the emission from the source is more centrally concentrated than this, it does not matter because Fig. 3 and Fig. 6 suggest that the cross-section of a source of our chosen $\Gamma$ (and hence $R_0$, if we do not vary $M_0$) to lensing from an isothermal sphere is approximately equal to that of a point source. It is unlikely that the emission is **less** centrally concentrated than this model suggests (the CO observations of Radford et al. (1993) show that the emission is only slightly extended). Figure 3 suggests that that significant amounts of the luminosity would need to come from regions approximately 20 kpc from the nucleus for the cross-sections to differ significantly.

(ii) Lens mass: we have assumed a constant lens mass. Our adopted values of $R_0$ and $\Gamma$ suggest a lens mass of $5 \times 10^{11}$ M$_\odot$. This mass was chosen for reasons given in Section 3, namely that it is the typical total mass of a field giant galaxy. However, recent studies of the field galaxy luminosity function suggest that dwarf galaxies might be very numerous (the Schechter $\alpha^*$ might be as steep as $-1.8$; Marzke et al. 1994). These dwarf galaxies are increasingly dark-matter dominated at lower luminosities (Kormendy 1990) so that a considerable fraction of the cosmological mass density in galaxies might reside in their halos. This would suggest a lower value of $M_0$ might be appropriate. This would increase $R_0$ and thus (see Fig. 3) cause $\mu_c$, the magnification at which the $\mu_e^{-2}$ part of the profile turns over to the $\mu_e^{-6}$ part, to decrease. This would then result a substantial decrease in $\frac{\rho(\mu_e)}{\rho(1)}$ for values of $\mu_e > \mu_c$.

(iii) Lens density: there is increasing dynamical evidence (e.g. Tully et al. 1993, White et al. 1993) that the cosmological density of matter that clumps on scales less than 20 Mpc is approximately 0.2. Also, this is close to the value of $\Omega$ that is found if the cosmological density that is in luminous matter (most of which is in giant galaxies, Faber & Gallagher 1979) is multiplied by a typical global (i.e. including dark matter) mass-to-light ratio for a giant galaxy (Trimble 1987). However, our choice of $\Omega_L = 0.2$ should be regarded as optimistic for the lensing case; it would be an overestimate if much of the dynamical mass is in dwarf galaxies or their tidally disrupted remnants. Furthermore, recent hydrodynamic simulations (Cen et al. 1994 and references therein) suggest that the value of $\Omega$ in collapsed objects may be substantially lower than 0.2.

In summary, the variation in $\frac{\rho(\mu_e)}{\rho(1)}$ resulting from changes in $\alpha$, $M_0$, and $\Omega_L$ is given by the following approximate relation:

$$\frac{\rho(\mu_e)}{\rho(1)} \approx 0.25 \Omega_L{}^{1.5} \mu_e{}^{-\alpha-2.0}, \qquad (11a)$$

if

$$\mu_e < \mu_c,$$

where

$$\mu_c \approx 10 \left(\frac{M_0}{5 \times 10^{11} M_\odot}\right)^{0.4} \qquad (11b)$$

For large magnifications ($\mu_e > 2\mu_c$), the corresponding relation is $\frac{\rho(\mu_e)}{\rho(1)} \approx 1.5 \Omega_L{}^{1.5} \mu_e{}^{-\alpha-6.0}$; such high magnifications are therefore ruled out unless $\phi(L)$ is very steep. Note also that if $M_0$ is small (i.e. small galaxies dominate the lens mass density), a very large value of $|\alpha|$ is required if the probability of F10214 being lensed is not to become negligible.

It should be noted that the statistical calculations presented here do not rigorously answer the question as to whether or not F10214 is a gravitational lens system. Instead, they give a probabilistic statement of how likely it is that F10214 is a lens with total magnification $\mu_e$, given a value of $\alpha$ and a particular cosmology (in the example presented here, a matter-dominated Friedmann cosmology). This statement, albeit a probabilistic one, has several implications for our understanding of both F10214 and the class of objects of which it is a hypothesized member, the ULIGs. These implications are discussed in the remainder of this section.

Figure 5 suggests that the optimal redshift for lensing with moderate magnifications ($\mu_e \sim 5$) is about 0.5. This is close to the redshift of the foreground galaxies observed by Elston et al. (1994), as inferred from luminosity and colour information. It is also close to the predicted lens redshift if Source 1 of Matthews et al. is to be interpreted as an arc if the lensing object is a large galaxy. Direct measurement of the redshift of the foreground galaxies of Elston et al. (1994) would be a useful diagnostic in determining whether or not F10214 is lensed; however, the peak in Fig. 5 is broad, so the lens redshift is not being strongly constrained by this model. If, on the other hand, the foreground galaxies are close to either the source or observer, they are ruled out as being candidate lenses. A more definitive test would be to perform high-resolution imaging polarimetry. The light from F10214 is known to be polarized over a substantial part of the spectrum (Lawrence et al. 1993, Januzzi et al. 1994). If it is subsequently found that the light from Source 1 of Matthews et al. (1994) is polarized but that from Source 2 is not, this might be regarded as evidence for F10214 being lensed. The most convincing evidence would be the detection of multiple components with the same polarization. Such high-resolution polarimetry is marginally feasible at present with the largest ground-based telescopes. This model also predicts that if F10214 is a lens, there ought to exist a population of ULIGs (both lensed and unlensed) at $z \sim 2$ that will be detected with the *Infrared Space Observatory* (ISO) satellite. If such a population is detected, we might be able (e.g. by searching for multiple components using radio interferometry) to measure the fraction of such objects that are lensed. This fraction is a strong function of the specific model parameters ($\alpha$, $\Omega_L$, $M_0$, etc.), and can therefore be used to constrain the particular models being investigated here. F10214 would then be an extreme example of such a population. Non-detection of such a population might also provide evidence supporting this model, as it would suggest that the luminosity function of ULIGs at $z \sim 2$ is very steep indeed. However, in this case there would exist the plausible alternative that F10214 is just undergoing an extremely short-lived high-luminosity phase (although at present there is no theoretical motivation for this).

The detection of a population of ULIGs at $z \sim 2$ with ISO might also offer a test of the cosmological parameters and the distribution of lensing material if we are able to accurately determine the fraction of these galaxies that are lensed. However, if the mean redshift of the sample is about 2, such a test is unlikely to produce stronger constraints than the existing studies of the lensing of the quasar population at even higher redshifts. Nevertheless, if ULIGs have an extremely steep intrinsic luminosity function at $z \sim 2$, then the effect of gravitational lensing on the luminosity function will be more marked than for QSOs at that redshift, and they still can be an important probe. Gravitational lensing may also give us the opportunity to study high redshift ULIGs that we would otherwise be unable to see. Morphological information may be distorted by the lensing, so that the above statement is particularly relevant to spectroscopic studies, for example at submillimetre wavelengths. If a lens, F10214 is an object whose CO properties are only known because of the lensing amplification. At least one other such object is known (the Cloverleaf quasar, Barvainis et al. 1994). The ISO satellite may reveal several more. If ULIGs do indeed have a steep luminosity function, our calculations suggest that most of the first examples discovered at high redshifts ought to be gravitational lenses.

If F10214 is to be interpreted as a gravitational lens, we might ask if this has any implications for settling the question as to whether its primary energy source is a starburst or an embedded quasar. The similarity of the SED to that of both Mrk 231 and Arp 220 might suggest that an embedded quasar is a possibility, as both of these objects are known to contain powerful embedded active galactic nuclei (see Lonsdale et al. 1994 for Arp 220 and Sanders et al. 1988 for Mrk 231). However, the X-ray measurements might rule this possibility out as argued by Lawrence et al. (1994), but these arguments would then also rule out Mrk 231 and Arp 220 as being predominantly quasar powered. On the other hand, the arguments of Lawrence et al. (1994) are weakened if there is significant uncertainty in the X-ray ($\sim 1$ keV) extinctions of these galaxies (Eales & Arnaud 1988). However, we recognize that drawing such conclusions based on the similarity of SEDs alone is dangerous, and do not offer a compelling solution to this question.

## 5. SUMMARY

In summary, the evidence for F10214 being a lens system with a significant magnification is mostly circumstantial and is described in Section 2 of this paper; the statistical considerations (which follow naturally from a Friedmann cosmology and a plausible distribution of matter in the Universe) argue (on a probabilistic basis) that such a scenario is physically reasonable. On statistical grounds alone, the probability that F10214 is a gravitational lens system with magnification $2 < \mu_e < 10$ is approximately 25%, given the KS95 luminosity function extrapolated to luminosities $\sim 10^{14}$ L$_\odot$ and redshift $z = 2.286$. This luminosity function is for a sample of ULIGs at $z \sim 0.3$ and with mean luminosity $10^{12.5}$ L$_\odot$; a steepening of the luminosity function at higher redshifts and/or at higher luminosities would result in a substantial increase in the probability of F10214 being lensed. Magnifications larger than 20 are ruled out by this model unless $\phi(L)$ is extremely steep. If F10214 is to be interepreted as a lens with magnification $2 < \mu_e < 10$, the bolometric luminosity of F10214 would then be $\sim 4 \times 10^{13}$ L$_\odot$. This would still make it the most luminous galaxy known, but it would then have escaped inclusion in the IRAS Faint Source Catalog (and hence Rowan-Robinson's 0.2 Jy sample) had it not been lensed.

We suggest the following three tests of this model:

(i) A determination of the spectroscopic redshift of the foreground objects found by Elston et al. (1994) might be useful. Whilst a redshift of $z \sim 0.5$ for this group would lend weight to a lens interpretation for F10214, a redshift that differs from this does not necessarily rule out the lens models because of the broad peak of Fig. 5. Only a redshift very close to the source or observer would rule out the possibility of their being lenses for F10214 (although of course this would not necessarily rule out the existence of other lenses).

(ii) High resolution imaging polarimetry might reveal morphologies that are consistent with a lensing geometry. For example, if multiple images with the same polarization are found this would be convincing evidence supporting the lensing case.

(iii) The detection with ISO of a population of ULIGs at $z \sim 2$ with a luminosity function that exhibits a high amplification bias at high luminosities. F10214 would then be a very luminous example of such a galaxy, but ISO should be able to detect others given its lower flux limit.

## ACKNOWLEDGMENTS

The author wishes to thank J. Goldader, D. Sanders, and J. Graham for helpful discussions, and for permission to quote results prior to publication, and A. Evans for a careful reading of the manuscript.

# REFERENCES

Barvainis R., Tacconi L., Antonucci R., Alloin D., Coleman P., 1994, Nat, 1994, 371, 586

Brown R. L., Vanden Bout P. A., 1991, AJ, 102, 1956

Cen R., Gnedin N. Y., Ostriker J. P., 1993, ApJ, 417, 387

Condon J. J., Broderick J. J., 1991, AJ, 102, 1663

David L. P., Jones C., Forman W., 1992, ApJ, 388, 82

Davis M., Huchra J., 1982, ApJ, 254, 437

Dyer C. C., Roeder R. C., 1973, ApJ, 180, L31

Dyer C. C., Shaver E. G., 1992, ApJ, 390, L5

Eales S. A., Arnaud K. A., 1988, ApJ, 324, 193

Elston R., McCarthy P. J., Eisenhardt P., Dickinson M., Spinrad H., Januzzi B. T., Mahoney P., 1994, AJ, 107, 910

Faber S. M., Gallagher J. S., 1979, ARA&A, 17, 135

Felten J. E., 1985, Comments Ap, 11, 53

Ferguson H. C., Sandage A., 1991, AJ, 101, 765

Fisher K. B., Strauss M. A., Davis M., Yahil A., Huchra J. P., 1992, ApJ, 389, 188

Fukugita M., Turner E. L., 1991, MNRAS, 253, 99

Giallongo E., Vagnetti F., 1992, ApJ, 396, 411

Graham J. R., Liu M. C., 1995, ApJL, submitted

Januzzi B. T., Elston R., Schmidt G. D., Smith P. S., Stockman H. S., 1994, ApJ, 429, L49

Kayser R., Refsdal S., 1988, A&A, 197, 63

Kayser R., Surdej J., Condon J. J., Kellermann K. I., Magain P., Remy M., Smette A., 1990, ApJ, 364, 15

Kim D. C., Sanders D. B., 1995, ApJ, submitted

Kofman L. A., Gnedin N. Y., Bahcall N. A., 1993, ApJ, 413, 1

Kormendy J., 1990, in Kron R. G., ed., The Edwin Hubble Centennial Symposium: The Evolution of the Universe of Galaxies. Astronomical Society of the Pacific, San Francisco, p. 33

Lawrence A. et al., 1993, MNRAS, 260, 28

Lawrence A., Rigopoulou D., Rowan-Robinson M., McMahon R. G., Broadhurst T., Lonsdale C. J., 1994, MNRAS, 266, L41

Lonsdale C. J., Diamond P. J, Smith H. E, Lonsdale C. J., 1994, Nat, 370, 117

Marshall H. L., 1985, ApJ, 299, 109

Marzke R. O, Geller M. J., Huchra J. P., Corwin H. G., 1994, preprint

Matthews K. et al., 1994, ApJ, 420, L13

Peacock J. A., 1982, MNRAS, 199, 987

Radford S. J. E., Brown R. L., Vanden Bout P. A. 1993, A&A, 271, L71

Rowan-Robinson M. et al., 1991, Nat, 351, 719

Rowan-Robinson M. et al., 1993, MNRAS, 261, 513

Sanders D. B., Soifer B. T., Elias J. H., Madore B. F., Matthews K., Neugebauer G., Scoville N. Z., 1988, ApJ, 325, 74

Saunders W., Rowan-Robinson M., Lawrence A., Efstathiou G., Kaiser N., Ellis R. S., Frenk C. S., 1990, MNRAS, 242, 318

Schechter P., 1976, ApJ, 203, 297

Schneider P., Ehlers J., Falco E. E., 1992, Gravitational Lenses. Springer-Verlag, New York

Solomon P. M., Downes D., Radford S. J. E., 1992a, ApJ, 398, L29

Solomon P. M., Downes D., Radford S. J. E., 1992b, ApJ, 387, L55

Solomon P. M., Radford S. J. E., Downes D., 1992, Nat, 356, 318

Trimble V. A., 1987, ARAA, 25, 425

Tully R. B., Shaya E. J., Peebles P. J. E., 1993, Institute for Astronomy preprint

Turner E. L., 1990, ApJ, 365, L43

Turner E. L, Ostriker J. P., Gott J. R., 1984, ApJ, 284, 1

White S. D. M., Navarro J. F., Evrard A. E., Frenk C. S., 1993, Nat, 366, 429